\documentclass[sigconf]{acmart}

\renewcommand\footnotetextcopyrightpermission[1]{}

\usepackage{booktabs}
\usepackage{float}
\usepackage{url}
\usepackage{xcolor,colortbl}
\usepackage{csquotes}
\usepackage[nolist,nohyperlinks]{acronym}

\usepackage{listings}
\colorlet{punct}{red!60!black}
\definecolor{background}{HTML}{EEEEEE}
\definecolor{delim}{RGB}{20,105,176}
\colorlet{numb}{magenta!60!black}
\lstdefinelanguage{json}{
	basicstyle=\normalfont\ttfamily,
	numbers=left,
	numberstyle=\scriptsize,
	stepnumber=1,
	numbersep=8pt,
	showstringspaces=false,
	breaklines=false,
	literate=
	*{0}{{{\color{numb}0}}}{1}
	{1}{{{\color{numb}1}}}{1}
	{2}{{{\color{numb}2}}}{1}
	{3}{{{\color{numb}3}}}{1}
	{4}{{{\color{numb}4}}}{1}
	{5}{{{\color{numb}5}}}{1}
	{6}{{{\color{numb}6}}}{1}
	{7}{{{\color{numb}7}}}{1}
	{8}{{{\color{numb}8}}}{1}
	{9}{{{\color{numb}9}}}{1}
	{:}{{{\color{punct}{:}}}}{1}
	{,}{{{\color{punct}{,}}}}{1}
	{\{}{{{\color{delim}{\{}}}}{1}
	{\}}{{{\color{delim}{\}}}}}{1}
	{[}{{{\color{delim}{[}}}}{1}
	{]}{{{\color{delim}{]}}}}{1},
}

\newcommand{\eg}{e.g.~}
\newcommand{\ie}{i.e.~}
\newcommand{\katti}{\textsc{Katti}~}

\begin{document}
\title{Dismantling Common Internet Services\\for Ad-Malware Detection}

\author{Florian Nettersheim}
\email{{firstname.lastname}@bsi.bund.de}
\affiliation{
	\institution{Federal Office\\for Information Security}
	\city{Bonn}
	\country{Germany}
}
\author{Stephan Arlt}
\email{{firstname.lastname}@bsi.bund.de}
\affiliation{
	\institution{Federal Office\\for Information Security}
	\city{Bonn}
	\country{Germany}
}
\author{Michael Rademacher}
\email{{firstname.lastname}@h-brs.de}
\affiliation{
	\institution{University of Applied Sciences Bonn-Rhein-Sieg}
	\city{Sankt Augustin}
	\country{Germany}
}

\begin{abstract}
Online advertising represents a main instrument for publishers to fund content on the World Wide Web.
Unfortunately, a significant number of online advertisements often accommodates potentially malicious content, such as cryptojacking hidden in web banners -- even on reputable websites.
In order to protect Internet users from such online threats, the thorough detection of ad-malware campaigns plays a crucial role for a safe Web.
Today, common Internet services like VirusTotal can label suspicious content based on feedback from contributors and from the entire Web community.
However, it is open to which extent ad-malware is actually taken into account and whether the results of these services are consistent. 

In this pre-study, we evaluate who defines ad-malware on the Internet.
In a first step, we crawl a vast set of websites and fetch all HTTP requests (particularly to online advertisements) within these websites.
Then we query these requests both against popular filtered DNS providers and VirusTotal.
The idea is to validate, how much content is labeled as a potential threat.

The results show that up to 0.47\% of the domains found during crawling are labeled as suspicious by DNS providers and up to 8.8\% by VirusTotal. Moreover, only about 0.7\% to 3.2\% of these domains are categorized as ad-malware.
The overall responses from the used Internet services paint a divergent picture:
All considered services have different understandings to the definition of suspicious content. 
Thus, we outline potential research efforts to the automated detection of ad-malware.
We further bring up the open question of a common definition of ad-malware to the Web community.
\end{abstract}




\maketitle
\pagestyle{plain}

\section{Introduction}
\label{sec:introduction}

Online advertising is a main instrument for publishers to fund content for their websites.
While most advertising is harmless, malvertising represents a growing threat~\cite{LiZXYW12,geoedgeReport}.  
This includes cryptojacking, phishing attacks, drive-by-downloads, and other doubtful content.
Malicious actors proceed with a high level of expertise and are thus be able to circumvent security mechanisms on the part of the advertising industry~\cite{geoedgeReport,malwarebytesReport}.
Note that ad-malware is clearly harmful to Internet users and must not be confused with other advertisements (\eg trackers) that may raise other concerns (\eg privacy~\cite{EnglehardtN16}).
For website publishers it means a challenging task~\cite{ZarrasKSHKV14} to secure their websites, as they usually do not have full control of the advertisements:
The interaction of numerous systems leads to the actual advertisement impression chosen from a real-time bidding system of the advertising network~\cite{rtbBook}.

Some studies suggest that the online advertising ecosystem is broken from a security and privacy perspective~\cite{cai_threats_2020}.
Thus, the thorough detection of ad-malware campaigns plays a crucial role of the future security, safety, and ultimately health of the Web.\cite{LiZXYW12}
The user's options for protection are limited.
On the one hand, the use of ad blocking technologies can prohibit the display of any online advertising and thus malvertising.
On the other hand, it is possible to use \ac{TI} services, such as filtered DNS providers, Google-Safe-Browsing or crowd-based approaches like \ac{VT}.
Users can enroll to such services based on their requirements.
However, when using these services, the justification whether some content is identified as malicious is not transparently discernible to users.
Furthermore, it is open to which extent especially ad-malware is actually taken into account and whether the results differ among these services.

In this short paper we conduct a pre-study on who defines ad-malware on common Internet services.
We select a vast set of URLs from the Tranco list~\cite{PochatGTKJ19} representing popular websites of the Internet.
Then, we apply \katti~\cite{katti} on the set of URLs to crawl and fetch all data transferred on the application layer.
This includes in particular all HTTP requests included in the websites (particularly such that display online advertisements).
Then, we query all domains extracted from these requests against three filtered DNS endpoints~\cite{cloudflare,quad9,cisco}.
We further query all domains against the threat intelligence service \ac{VT}~\cite{vt}.
All services return information, whether a domain is labeled as a potential threat.
In a final step, we compare the results coming from these services.

The overall responses from the used Internet services paint a divergent picture:
All considered services have different understandings to the definition of suspicious content.
We thus conclude with a open question for the Web community, namely: \enquote{What could be a common definition of ad-malware?}

The next section presents our current approach to ad-malware detection.
In Section~\ref{sec:approach} we perform a brief evaluation using two different sources (\ie filtered DNS providers and \ac{VT}) of threat intelligence.
This includes a discussion of the results obtained from our evaluation.
We provide related work in Section~\ref{sec:relatedwork} and outline future work in Section~\ref{sec:conclusion}.

\section{Approach}
\label{sec:approach}

\begin{figure}[ht]
	\caption{Overview of our current approach to ad-malware detection.}
	\centering
	\includegraphics[width=\columnwidth]{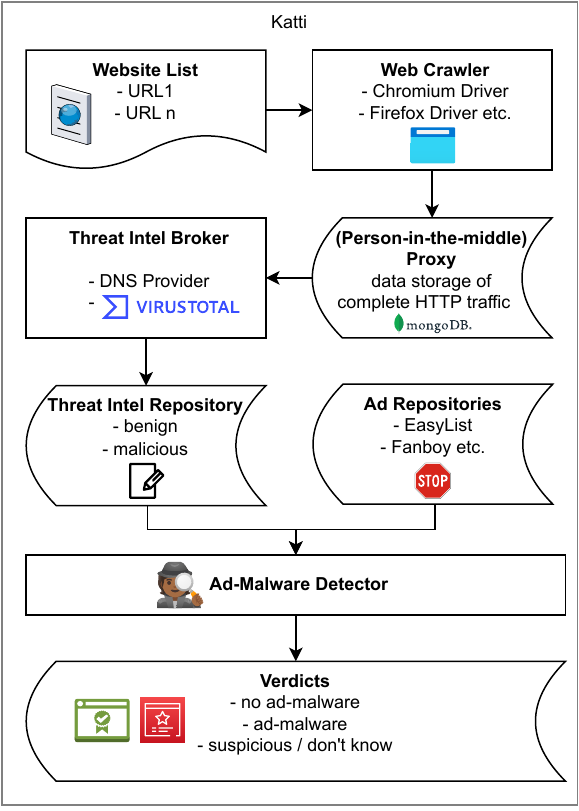}
	\label{fig:approach}
\end{figure}

In our approach, we extend our tool~\katti~\cite{katti} for the detection of ad-malware, which is visualized in Figure~\ref{fig:approach}.
In a first step, we take a list of websites (\eg the Tranco list~\cite{PochatGTKJ19}) and choose a web browser (\eg Chrome) to crawl all URLs of the list.
Note that \katti~employs real web browsers for crawling websites, which allows us to utilize more browsers such as Firefox (or even TOR) to consider cloaking~\cite{InvernizziTKCPB16}.
A person-in-the-middle HTTP proxy records all traffic passing through the application layer and stores it in the data storage.
We save all URLs visited in the crawling process, especially the HTTP requests called within websites, that is, URLs to online advertisements.

In a second step, our \emph{Threat Intel Broker} takes all HTTP requests found in the data storage and performs queries against multiple \ac{TI} services.
In this pre-study we confine the approach on public, filtered DNS providers and \ac{VT} \cite{vt}.
DNS servers translate domain names into IP addresses and are further able to block malicious domains.
\ac{VT} is an established service for \ac{TI} within the cybersecurity community and is often used for data labeling or system evaluation.
It maintains a vast dataset of potentially suspicious content based on feedback from multiple contributors~\cite{vt}.
The results delivered from these services are stored in our \emph{Threat Intel Repository}.
More precisely, for each HTTP request from our crawling process, our approach stores knowledge especially whether the corresponding data item is potentially benign or malicious.
Note that some services may return no result (or \enquote{don't know}), which we discuss in Section~\ref{sec:evaluation}.

In a third step, our \emph{Ad-Malware Detector} takes as input the \ac{TI} repository and an \emph{Ad Repository}.
The detector is responsible for deciding whether an HTTP request is related to online advertising.
This allows us to filter out all online advertisements among the content within the \ac{TI} repository (which may also contain non ad-related content).
By combining these two information sources from \ac{TI} services and from ad repositories, our approach is able to identify ad-malware.
In this pre-study, we leave our approach fully automated.
However, in a future work, deep manual inspection in AI augmentation seems a promising line of research and is adaptable in \katti.
Finally, the detector returns a verdict for each analyzed HTTP request, that is, online advertisement.

\section{Evaluation and Discussion}
\label{sec:evaluation}

This section provides a brief evaluation of research question to our approach.
For this purpose, we validated a total of 1,206,803 domains with the help of different DNS providers and \ac{VT}.
The domains originate from crawls of websites that have already been carried out.
We have deliberately used domains of the Tranco list~\cite{PochatGTKJ19} and domains from past crawls~\cite{katti}, which are possibly associated with the display of online advertising.
The crawls were carried out at the end of 2023 and were generally based on the Tranco 1 million list. 

\textbf{RQ1: How many domains are blocked from DNS providers?}\\
In this research question, we query all domains extracted from HTTP requests in our data storage against the DNS endpoints of Cisco~\cite{cisco}, Quad9~\cite{quad9}, and Cloudflare\cite{cloudflare}.
All three services offer so-called filtered endpoints.
When DNS requests are sent to the filtered endpoints, the response indicates whether the domain is blocked.
\katti utilizes and instruments the tool \textit{dig} for all DNS handling, and all queries of our evaluation are performed in a narrow time window from 17/12/2023 to 18/12/2023.
This ensures the comparability of the results, as DNS zones (\eg A records) may change frequently.
Note that we do not consider DNS providers that block all online advertisements (ad blockers) as we aim to find malicious ad impressions among all (potentially benign) advertisements.
To the best of our knowledge, we are not aware of any previous work that explicitly uses DNS providers for labeling HTTP resources.

\begin{figure}
	\caption{Blocked domains from different DNS providers.}
	\centering
	\includegraphics[width=.5\columnwidth]{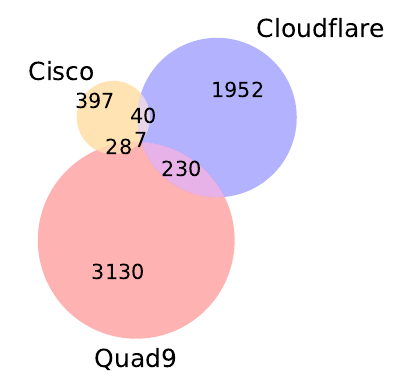}
	\label{fig:venn3}
\end{figure}

Our naive assumption for RQ1 is that all providers label (\ie block) most of the same domains as malicious resulting in a significant intersection.
Figure~\ref{fig:venn3} shows the results of our analysis.

The results show that Quad9 labels 3,395 domains (\ie 0.28\%) as malicious, Cisco labels only 472 domains (\ie 0.03\%) as malicious and Cloudflare labels 2,229 domains (\ie 0.18\%) as malicious. 5,784 domains (\ie 0.47\%) are labeled as malicious by at least one DNS provider.
Quad9 and Cloudflare have the highest correlation with 230 domains.
With only 28 domains, Quad9 and Cisco have the lowest correlation.
Interestingly, only 7 domains from more than 1.2 millions domains are blocked from all DNS servers.
Hence, our above-mentioned assumption is not confirmed, such that DNS providers do actually have different understandings which kind of domains are labeled as malicious.

\begin{figure}[ht]
	\centering
	\includegraphics[width=\columnwidth]{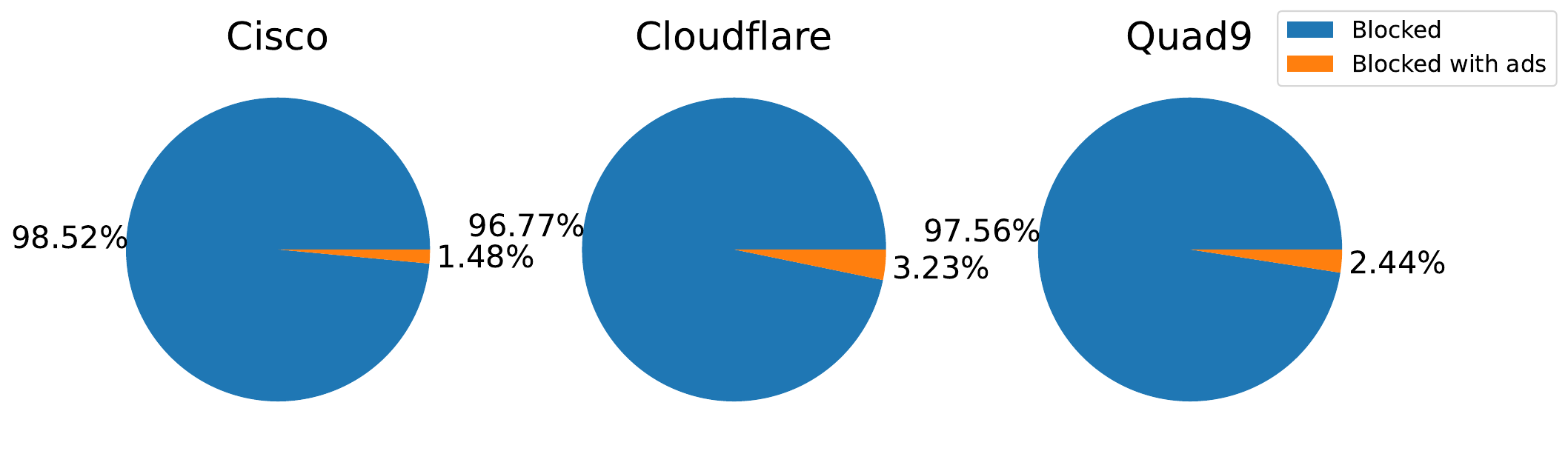}
	\caption{Blocked domains by DNS provider filtered by online advertisements.}
	\label{fig:ad_stats_pie}
\end{figure}

\textbf{RQ2: How many of the blocked domains are ad-malware?}\\
In this research question, we take the set of blocked domains from our RQ1 and assess, how many domains represent online advertisements, and thus, potentially ad-malware.
To decide whether a domain is associated with advertising, we checked it against multiple advertising filter lists from the Pi-hole project~\cite{pihole}.
If a domain has a positive match on one of the filter lists and is blocked by a DNS provider, we assume that the domain is connected to ad-malware.
Here, our naive assumption is that we obtain a decent number of ad-malware domains, since ad-malware may be a significant number among all malicious domains (\eg phishing domains).
Figure~\ref{fig:ad_stats_pie} shows the results of our analysis.

The results show that the set of blocked domains from Cloudflare contains most ad-malware domains (\ie 3.23\%), whereas Cisco considers only 1.48\% of the domains.
Hence, our above-mentioned assumption is also not confirmed as the number of ad-malware domains is significantly low. We will discuss both results in the next section.

\textbf{Discussion of RQ1 and RQ2}\\
In RQ1 we learned that the results are divergent.
We assume that DNS providers may have different understandings and policies which domains should be labeled as malicious.
However, information on blocking criteria is largely intransparent to users.
Since DNS represents a fundamental Internet protocol and service, DNS providers certainly act carefully with blocking of domains to prevent overblocking, and thus, access to websites.
Furthermore, most DNS providers certainly do not want to be liable for censorship.

In RQ2 we further observed that only a very small fraction of blocked domains are related to ad-malware.
As mentioned above, DNS providers may act carefully when blocking domains.
In the context of online advertisements, this scenario becomes more realistic in the following example:
\begin{itemize}
\item[] \url{https://ads.example.com/?display=benign}
\item[] \url{https://ads.example.com/?display=malicious}
\end{itemize}

Assume the server ads.example.org displays both benign and potentially malicious ads.
Blocking the entire domain ads.example.org leads in also blocking all ads from this server.
Thus, a deeper analysis, which takes into account the entire query string is needed and must be addressed in a future work.

\textbf{RQ3: How many domains are blocked from \ac{VT}?}\\
In this research question, we query all domains found\footnote{\ac{VT} did not provide any reports for a number of 37,141 domains, meaning that these domains were unknown to \ac{VT} at the time of the scan. We assume that the proportion of 3\% of the domains analyzed has no influence on our evaluation.} in our crawling efforts against \ac{VT}.
More precisely, we used the corresponding data enrichment endpoint\footnote{https://docs.virustotal.com/reference/domain-info} to check the individual domains.
The result for each domain is a so-called report \cite{vt}.
A report shows, among other things, how many \ac{VT} partners have categorized the domain as:

\begin{enumerate}
	\item \textit{harmless}: Partner thinks the domain is harmless.
	\item \textit{undetected}: Partner has no opinion about this domain.
	\item \textit{suspicious}: Partner thinks the domain is suspicious.
	\item \textit{malicious}: Partner thinks the domain is malicious.
\end{enumerate}

We aggregated the results for \enquote{suspicious} and \enquote{malicious} since we are interested in a potential threat to the users.
Overall, the number of domains where \textit{at least one partner} flags the domain as a potential threat is \textbf{8.8\%.} 
In addition, similar to RQ2, we are interested in the share of online advertisements from these potential threats. 
Only \textbf{0.71\%} from the potential threats are identified as advertisements.

\begin{figure}[ht]
	\caption{Opinion of different \ac{VT} partners if a certain domain is considered harmless or a potential threat (malicious and suspicious).}
	\centering
	\includegraphics[width=\columnwidth]{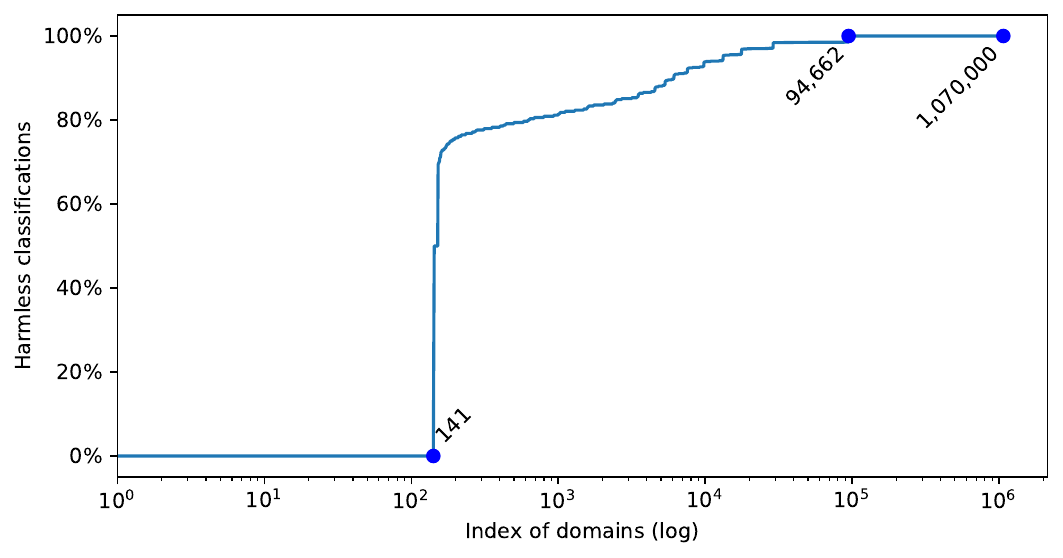}
	\label{fig:vt_partners_cdf}
\end{figure}

\textbf{RQ4: How consistent are the results from \ac{VT} partners?}\\
\ac{VT} works with a variety of different contributors (\eg \enquote{Google Safebrowsing}, \enquote{Fortinet}, \enquote{Avira}, \ldots) to provide a differentiated opinion if a certain domain is a potential threat.
Similar to the second part of RQ1, we are interested if the opinions of the different partners vary, which is visualized in Figure~\ref{fig:vt_partners_cdf}.
For all domains, it is shown how many (given in \%) partners evaluated the domain as a potential threat.
For 141 domains, all partners agreed that the domain is a potential threat (either malicious or suspicious).
For 975,338 (1,070,000-94,662) domains, all partners agreed that the domain is harmless.
Inbetween, for overall 94,521 (94,662-141) domains, the opinion differs.
However, as Figure~\ref{fig:vt_partners_cdf} reveals, the disagreement is skewed and not uniformly distributed.
There are many domains where the majority of partners $>80\%$ label the domain as harmless while certain partners see a potential threat.

\textbf{Discussion of RQ3 and RQ4}\\
Compared to the DNS providers, \ac{VT} flags significantly more domains as a potential threat to  users (0.47\% vs.\ 8.8\%).
In addition, there exists varying opinions among the different VT partners what is considered harmless and a potential threat to users. 
Additional research can evaluate dependencies among the different partners, for example, if certain partners differ significantly from others.

\section{Related Work}
\label{sec:relatedwork}

The challenge of interpretation and divergence of results when using \ac{TI} services is a well-known issue.
\ac{VT} is increasingly in the spotlight due to its role within the research community.
Peng et al.~\cite{PengYS019} shows that \ac{VT} scanners have similar inconsistencies in categorizing URLs as malicious and benign.
The authors also show that some \ac{VT} scanners are more correct than others, which requires a strategy for labeling such URLs that does not treat all scanners equally.

Salem et al.~\cite{SalemBP21} presented Maat, an approach designed to address the inconsistency in \ac{VT}'s scan reports.
Maat is a systematic method for generating ML-based labeling strategies based on the current scan results provided by \ac{VT}.
Furthermore, a critical review of \ac{VT} and its use within research per se takes place.~\cite{SalemBP21}

Hurier et al.~\cite{HurierABKT16} also looks at the labeling of the various \ac{VT} partners.
However, their focus was on files.
Nevertheless, a lack of consistency in labels was observed for the same file.

\section{Conclusion and Future Work}
\label{sec:conclusion}

When detecting ad-malware, it is important to have scalable and meaningful detection mechanisms for malicious web resources.
This is why \ac{TI} services such as \ac{VT} are essential for the evaluation of web resources such as domains.
They make it easy to participate in the knowledge of leading cybersecurity services.
The results of our pre-study have shown that we need to develop approaches for the interpretation and evaluation of the results of such \ac{TI} services for future work.
One promising approach is \textit{Maat} from Salem et al.~\cite{SalemBP21}.
The use of filtered DNS endpoints is a comparatively simple protective measure for Internet users.
Usually, only the DNS resolver in the home internet router needs to be changed.
Our results have shown that the DNS providers have a high degree of divergence in their results.
A study is planned to examine the effectiveness of such DNS servers in checking malicious domains (\eg phishing domains or malware domains).
An elementary part of ad-malware detection is to identify web resources that were involved in the delivery of ad impressions.
In the future, it is planned to use a more sophisticated approach for this.
One possibility is, for example, the instrumentation of the adblock engine of the Brave browser\footnote{https://github.com/brave/adblock-rust}.
In general, the varying options of \ac{TI} services and DNS providers show the need for a more general and transparent definition what is considered as malicious in the World Wide Web.

\begin{acronym}[]
\acro{3GPP}{3rd Generation Partnership Project}
\acro{AC}{Access Category}
\acro{ACK}{acknowledgement}
\acro{AI}{Abstract Interface}
\acro{AIFS}{Arbitration Interframe Space}
\acro{AODV}{Ad hoc On-Demand Distance Vector Routing Protocol}
\acro{AP}{Access Point}
\acro{API}{application programming interface}
\acro{ARP}{Address Resolution Protocol}
\acro{ARQ}{Automatic Repeat reQuest}
\acro{AS}{Autonomous System}
\acro{ASCII}{American Standard Code for Information Interchange}
\acro{ATIS}{Alliance for Telecommunications Industry Solutions}
\acro{ATM}{Asynchronous Transfer Mode}
\acro{B.A.T.M.A.N.}{Better Approach to MANET}
\acro{BER}{Bit Error Rate}
\acro{BFS-CA}{Breadth First Search Channel Assignment}
\acro{BGP}{Border Gateway Protocol}
\acro{BPSK}{Binary Phase-Shift Keying}
\acro{BRA}{Bidrectional Routing Abstraction}
\acro{BS}{Base station}
\acro{BSSID}{Basic Service Set Identification}
\acro{BTS}{Base Transceiver Station}
\acro{CA}{Channel Assignment}
\acro{CAPEX}{capital expenditure}
\acro{CAPWAP}{Control And Provisioning of Wireless Access Points}
\acro{CARD}{Channel Assignment with Route Discovery}
\acro{CAS}{Channel Assignment Server}
\acro{CCA}{Clear Channel Assessment}
\acro{CDMA}{Code Division Multiple Access}
\acro{CF}{CompactFlash}
\acro{CIDR}{Classless Inter-Domain Routing}
\acro{CLICA}{Connected Low Interference Channel Assignment}
\acro{CLI}{Command Line Interface}
\acro{COTS}{Commercial Off-the-Shelf}
\acro{CPE}{Customer Premises Equipment}
\acro{CPU}{Central Processing Unit}
\acro{CRAHN}{Cognitive Radio Ad-Hoc Network}
\acro{CRCN}{Cognitive Radio Cellular Network}
\acro{CR}{Cognitive Radio}
\acro{CR-LDP}{Constraint-based Routing Label Distribution Protocol}
\acro{CRN}{Cognitive Radio Network}
\acro{CRSN}{Cognitive Radio Sensor Network}
\acro{CRVN}{Cognitive Radio Vehicular Network}
\acro{CSMA/CA}{Carrier Sense Multiple Access/Collision Avoidance}
\acro{CSMA}{Carrier Sense Multiple Access}
\acro{CSMA/CD}{Carrier Sense Multiple Access/Collision Detection}
\acro{CTA}{Centralized Tabu-based Algorithm}
\acro{CW}{Contention Window}
\acro{CWLAN}{Cognitive Wireless Local Area Network}
\acro{CWMN}{Cognitive Wireless Mesh Network}
\acro{DAD}{Duplicate Address Detection}
\acro{DCF}{Distributed Coordination Function}
\acro{DCiE}{Data Center Infrastructure Efficiency}
\acro{DDS}{Direct digital synthesizer}
\acro{DFS}{Dynamic Frequency Selection}
\acro{DGA}{Distributed Greedy Algorithm}
\acro{DHCP}{Dynamic Host Configuration Protocol}
\acro{DIFS}{Distributed Interframe Space}
\acro{DMesh}{Directional Mesh}
\acro{D-MICA}{Distributed Minimum Interference Channel Assignment}
\acro{DR}{Designated Router}
\acro{DSA}{Dynamic Spectrum Allocation}
\acro{DSLAM}{Digital Subscriber Line Access Multiplexer}
\acro{DSL}{Digital Subscriber Line}
\acro{DSR}{Dynamic Source Routing Protocol}
\acro{DSSS}{Direct-Sequence Spread Spectrum}
\acro{DTCP}{Dynamic Tunnel Configuration Protocol}
\acro{DVB}{Digital Video Broadcast}
\acro{DVB-H}{Digital Video Broadcast - Handheld}
\acro{DVB-RCS}{Digital Video Broadcast - Return Channel Satellite}
\acro{DVB-S2}{Digital Video Broadcast - Satellite - Second Generation}
\acro{DVB-S}{Digital Video Broadcast - Satellite}
\acro{DVB-SH}{Digital Video Broadcast - Satellite services to Handhelds}
\acro{DVB-T2}{Digital Video Broadcast - Second Generation Terrestrial}
\acro{DVB-T}{Digital Video Broadcast - Terrestrial}
\acro{E2CARA-TD}{Energy Efficient Channel Assignment and Routing Algorithm – Traffic Demands}
\acro{ECN}{Explicit Congestion Notification}
\acro{ECDF}{Empirical Cumulative Distribution Function}
\acro{EDCA}{Enhanced Distributed Coordination Access}
\acro{EDCF}{Enhanced Distributed Coordination Function}
\acro{EGP}{Exterior Gateway Protocol}
\acro{EICA}{External Interference-Aware Channel Assignment}
\acro{EIFS}{Extended Interframe Space}
\acro{EIGRP}{Enhanced Interior Gateway Routing Protocol}
\acro{EIRP}{Equivalent Isotropically Radiated Power}
\acro{EPI}{energy proportionality index}
\acro{ERO}{Explicit Route Object}
\acro{ETSI}{European Telecommunications Standards Institute}
\acro{ETT}{Expected Transmission Time}
\acro{ETX}{Expected Transmission Counts}
\acro{FCC}{Federal Communications Commission}
\acro{FCS}{Frame Check Sequence}
\acro{FDMA}{Frequency Division Multiple Access}
\acro{FEC}{Forward Error Correction}
\acro{FIFO}{First-In-First-Out}
\acro{FLOPS}{Floating-Point Operations Per Second}
\acro{FRR}{Fast Reroute}
\acro{FSL}{Free-Space Loss}
\acro{FSPL}{Free-Space Path Loss}
\acro{GAN}{Generic Access Network}
\acro{GDP}{Gross Domestic Product}
\acro{GEO}{Geosynchronous Earth Orbit}
\acro{GMPLS}{Generalized Multiprotocol Label Switching}
\acro{GPS}{Global Positioning System}
\acro{GRE}{Generic Routing Encapsulation}
\acro{GSE}{Generic Stream Encapsulation}
\acro{GSM}{Global System for Mobile Communications}
\acro{GW}{Gateway}
\acro{HAP}{High-Altitude Platform}
\acro{HCCA}{HCF controlled channel access}
\acro{HCF}{Hybrid Coordination Function}
\acro{HLR}{Home Location Register}
\acro{HOL}{Head-of-line}
\acro{HOLSR}{Hieracical Optimised Link State Routing}
\acro{HPC}{hardware performance counters}
\acro{HSLS}{Hazy-Sighted Link State Routing Protocol}
\acro{HWMP}{Hybrid Wireless Mesh Protocol}
\acro{IAX2}{Inter-Asterisk eXchange Version 2}
\acro{IBSS}{Independent Basic Service Set}
\acro{ICMP}{Internet Control Message Protocol}
\acro{ICT}{Information and Communication Technologie}
\acro{IEEE}{Institute of Electrical and Electronics Engineers}
\acro{IE}{Information Element}
\acro{IETF}{Internet Engineering Task Force}
\acro{IETF}{The Internet Engineering Task Force}
\acro{IFS}{Interframe Space}
\acro{ITU}{International Telecommunication Union}
\acro{IGP}{Interior Gateway Protocol}
\acro{IGRP}{Interior Gateway Routing Protocol}
\acro{ILP}{Integer Linear Programming}
\acro{ILS}{Iterated Local Search}
\acro{IPFIX}{IP Flow Information Export}
\acro{IP}{Internet Protocol}
\acro{IPv4}{Internet Protocol}
\acro{IPv6}{Internet Protocol, Version 6}
\acro{ISI}{Inter-symbol interference}
\acro{IS-IS}{Intermediate system to intermediate system}
\acro{ISM}{Industrial, Scientific and Medical}
\acro{ISP}{Internet Service Provider}
\acro{LAA}{Licensed-Assisted Access}
\acro{LDC}{Least Developed Countries}
\acro{LA-CA}{Load-Aware Channel Assignment}
\acro{LCOS}{LANCOM Operating System}
\acro{LDP}{Label Distribution Protocol}
\acro{Ld}{Log-distance}
\acro{LDPL}{Log-distance path loss}
\acro{LEO}{Low Earth Orbit}
\acro{LER}{Label Edge Router}
\acro{LGI}{Long Guard Interval}
\acro{LLTM}{Link Layer Tunneling Mechanism}
\acro{LMA}{Local Mobility Anchor}
\acro{LMP}{Link Management Protocol}
\acro{LoS}{Line of Sight}
\acro{LOS}{Line of Sight}
\acro{LQF}{Longest-Queue-First}
\acro{LS}{Link State}
\acro{LSP}{Label-Switched Path}
\acro{LSR}{Label-Switched Router}
\acro{LST}{Link-State-Table}
\acro{LTE}{Long Term Evolution}
\acro{LTE-M}{\ac{LTE} Machine Type Communication}
\acro{LWAPP}{Lightweight Access Point Protocol}
\acro{MAC}{Media Access Control}
\acro{MAG}{Mobile Access Gateway}
\acro{MANET}{Mobile Adhoc Network}
\acro{MBMS}{Multimedia Broadcast Multicast Service}
\acro{MCG}{multi-conflict graph}
\acro{MCI-CA}{Matroid Cardinality Intersection Channel Assignment}
\acro{MCS}{Modulation and Coding Scheme}
\acro{MDR}{MANET Designated Router}
\acro{MEO}{Medium Earth Orbit}
\acro{MICS}{Media Independent Command Service}
\acro{MIES}{Media Independent Event Service}
\acro{MIHF}{Media Independent Handover Function}
\acro{MIHF++}{Media Independent Handover Function++}
\acro{MIH}{Media Independent Handover}
\acro{MIIS}{Media Independent Information Service}
\acro{MILP}{Mixed Integer Linear Programming}
\acro{MIMF}{Media Independent Messaging Function}
\acro{MIMO}{Multiple Input Multiple Output}
\acro{MNO}{Mobile Network Operator}
\acro{MIMS}{Media Independent Messaging Service}
\acro{MIPS}{Million Instruction Per Second}
\acro{MMF}{Mobility Management Function}
\acro{MN}{Mesh Node}
\acro{MonF}{Monitoring Function}
\acro{MPDU}{MAC Protocol Data Unit}
\acro{MPEG}{Moving Picture Experts Group}
\acro{MPE}{Multi Protocol Encapsulation}
\acro{MPLCG}{Multi-Point Link Conflict Graph}
\acro{MPLS}{Multiprotocol Label Switching}
\acro{MPLS}{Multi Protocol Label Switching}
\acro{MPLS-TE}{Multi Protocol Label Switching - Traffic Engineering}
\acro{MP}{Merge Point}
\acro{MPR}{Multipoint Relay}
\acro{MR-MC WMN}{Multi-Radio Multi-Channel Wireless Mesh Network}
\acro{MSC}{Mobile-services Switching Centre}
\acro{MSDU}{MAC Service Data Unit}
\acro{MSTP}{Mobility Services Transport Protocol}
\acro{MT}{Mobile Terminal}
\acro{MTU}{Maximum Transmission Unit}
\acro{NAV}{Network Allocation Vector}
\acro{ns-3}{network simulator 3}
\acro{NBMA}{Non-broadcast Multiple Access}
\acro{NetEMU}{Network Emulator}
\acro{NLOS}{None Line of Sight}
\acro{NMEA}{National Marine Electronics Association}
\acro{NPC}{Normalized Power Consumption}
\acro{NP}{Nondeterministic Polynomial Time}
\acro{NSIS}{Next Steps in Signaling}
\acro{NTP}{Network Time Protocol Unit}
\acro{OFDMA}{Orthogonal Frequency Division Multiple Access}
\acro{OFDM}{Orthogonal Frequency Division Multiplex}
\acro{OLSR}{Optimized Link State Routing}
\acro{OPEX}{operational expenditure}
\acro{OSA}{Opportunistic Spectrum Access}
\acro{OSI}{Open Systems Interconnection}
\acro{OSPF}{Open Shortest Path First}
\acro{OSPF-TE}{Open Shortest Path First - Traffic Engineering}
\acro{OVS}{Open vSwitch}
\acro{P2MP}{Point To Multipoint}
\acro{P2P}{Point To Point}
\acro{PA}{Power amplifier}
\acro{PCE}{Path Computation Element}
\acro{PCEP}{Path Computation Element Protocol}
\acro{PCF}{Path Computation Function}
\acro{PCF}{Point Coordination Function}
\acro{PDR}{Packet Delivery Ratio}
\acro{PDV}{Packet Delay Variation}
\acro{PER}{Packet Error Rate}
\acro{PLCP}{Physical Layer Convergence Protocol}
\acro{PLL}{Phase-Locked Loop}
\acro{PL}{path loss}
\acro{PLR}{Point of Local Repair}
\acro{PMIP}{Proxy Mobile IP}
\acro{PoE}{Power over Ethernet}
\acro{PPDU}{Physical Protocol Data Unit}
\acro{PPP}{Precise Point Positioning}
\acro{PSTN}{Public Switched Telephone Network}
\acro{PTP}{Precision Time Protocol}
\acro{PUE}{Power Usage Effectiveness}
\acro{PU}{Primary User}
\acro{QAM}{Quadrature amplitude modulation}
\acro{QoS}{Quality of Service}
\acro{RAND}{Random Channel Assignment}
\acro{RERR}{Route Error}
\acro{RFC}{Request for Comments}
\acro{RIPng}{Routing Information Protocol next generation}
\acro{RIP}{Routing Information Protocol}
\acro{RPC}{Remote Procedure Call}
\acro{RPP}{Received Packet Power}
\acro{RREP}{Route Reply}
\acro{RREQ}{Route Request}
\acro{RSSI}{Received Signal Strength Indication}
\acro{RSS}{Received Signal Strength}
\acro{RSVP}{Resource ReSerVation Protocol}
\acro{RSVP-TE}{Resource ReSerVation Protocol - Traffic Engineering}
\acro{RTK}{Real Time Kinematic}
\acro{RTP}{Real-time Transport Protocol}
\acro{RTS}{Ready-To-Send}
\acro{RTT}{Round Trip Time}
\acro{SAA}{Stateless Address Autoconfiguration}
\acro{SAPOS}{Satellitenpositionierungsdienst der deutschen Landesvermessung}
\acro{SAP}{Service Access Point}
\acro{SBC}{Single-Board Computer}
\acro{SBM}{Subnetwork Bandwidth Manager}
\acro{SBR}{System zur Bestimmung des Richtungsfehlers}
\acro{SC-FDMA}{Single Carrier Frequency Division Multiple Access}
\acro{SDMA}{Space-division multiple access}
\acro{SDN}{Software Defined Networking}
\acro{SDR}{Software Defined Radio}
\acro{SDWN}{Software Defined Wireless Networks}
\acro{SENF}{Simple and Extensible Network Framework}
\acro{SGI}{Short Guard Intervall}
\acro{SIFS}{Short Interframe Space}
\acro{SINR}{Signal-to-Noise-plus-Interference Ratio}
\acro{SIP}{Session Initiation Protocol}
\acro{SISO}{Single Input Single Output}
\acro{SNR}{signal-to-noise ratio}
\acro{SONET}{Synchronous Optical Networking}
\acro{SPoF}{Single Point of Failure}
\acro{SSID}{Service Set Identifier}
\acro{STP}{Spanning Tree Protocol}
\acro{SU}{Secondary User}
\acro{TCP}{Transmission Control Protocol}
\acro{TPC}{Transmission Power Control}
\acro{TC}{Topology Control}
\acro{TDMA}{Time Division Multiple Access}
\acro{TI}{Threat Intelligence}
\acro{TEEER}{Telecommunications Equipment Energy Efficiency Rating}
\acro{TEER}{Telecommunications Energy Efficiency Ratio}
\acro{TE}{Traffic Engineering}
\acro{TIM}{Technology Independend Monitoring}
\acro{TLV}{Type-Length-Value}
\acro{TORA}{Temporally-Ordered Routing Algorithm}
\acro{ToS}{Type of Service}
\acro{TSFT}{Time Synchronization Function Timer}
\acro{TTL}{Time to live}
\acro{TVWS}{TV White Space}
\acro{TXOP}{Transmit opportunity}
\acro{UAV}{Unmanned Aerial Vehicle}
\acro{UDLR}{Unidirectional Link Routing}
\acro{UDL}{Unidirectional Link}
\acro{UDP}{User Datagram Protocol}
\acro{UDT}{Unidirectional Technology}
\acro{UE}{User Equipment}
\acro{UHF}{Ultra High Frequency}
\acro{UMA}{Unlicensed Mobile Access}
\acro{UMTS}{Universal Mobile Telecommunications System}
\acro{U-NII}{Unlicensed National Information Infrastructure}
\acro{UPS}{Uninterruptible Power Supply}
\acro{UP}{User Priorities}
\acro{USB}{Univeral Serial Bus}
\acro{USO}{Universal Service Obligation}
\acro{USRP}{Universal Software Radio Peripheral}
\acro{VCO}{Voltage-Controlled Oscillator}
\acro{VoIP}{Voice-over-IP}
\acro{VT}{VirusTotal}
\acro{VPN}{Virtual Prvate Network}
\acro{WBN}{Coordinated Wireless Backhaul Network}
\acro{WDS}{Wireless Distribution System}
\acro{WiBACK}{Wireless Back-Haul}
\acro{Wi-Fi}{Wireless Fidelity}
\acro{WiLD}{WiFi-based Long Distance}
\acro{WiMAX}{Worldwide Interoperability for Microwave Access}
\acro{WISPA}{Wireless Internet Service Provider Assocication}
\acro{WISP}{Wireless Internet Service Provider}
\acro{WLAN}{Wireless Local Area Network}
\acro{WLC}{Wireless LAN Controller}
\acro{WMN}{Wireless Mesh Network}
\acro{WMN}{Wireless Mesh Network}
\acro{wmSDN}{Wireless Mesh Software Defined Network}
\acro{WNIC}{Wireless Network Interface Controller}
\acro{WN}{WiBACK Node}
\acro{WRAN}{Wireless Regional Area Network}
\acro{WSN}{Wireless Sensor Network}
\acro{ZigBee}{ZigBee Alliance IEEE 802.15.4}
\acro{ZPR}{Zone Routing Protocol}

\acro{ABP}{Activation by Personalization}
\acro{ADR}{Adaptive Data Rate}
\acro{IoT}{Internet of Things}
\acro{LPWAN}{low-power wide-area network}
\acro{LoRaWAN}{Long Range Wide Area Network}
\acro{GPRS}{General Packet Radio Service}
\acro{CEP}{Circular Error Probable}
\acro{EIRP}{equivalent isotropically radiated power}
\acro{ITM}{Longley-Rice Irregular Terrain Model}
\acro{SF}{Spreading Factor}
\end{acronym}

\bibliographystyle{ACM-Reference-Format}
\bibliography{bibliography}

\end{document}